# RESEARCH ON CRO'S DILEMMA IN SAPIENS CHAIN: A GAME THEORY METHOD

Jinyu Shi[1], Zhongru Wang[1,2], Qiang Ruan[3], Yue Wu[1] and Binxing Fang[1]

[1]Key Laboratory of Trustworthy Distributed Computing and Service (BUPT),
Ministry of Education, Beijing, China
{shijinyu,wangzhongru, wuyue0530,fangbx}@bupt.edu.cn
[2]Zhejiang Lab, Hangzhou, China
[3]Beijing DigApis Technology Co., Ltd, Beijing, China
ruanqiang@digapis.cn

*ABSTRACT*

*In recent years, blockchain-based techniques have been widely used in cybersecurity, owing to the decentralization, anonymity, credibility and not be tampered properties of the blockchain. As one of the decentralized framework, Sapiens Chain was proposed to protect cybersecurity by scheduling the computational resources dynamically, which were owned by Computational Resources Owners (CROs). However, when CROs in the same pool attack each other, all CROs will earn less. In this paper, we tackle the problem of prisoner's dilemma from the perspective of CROs. We first define a game that a CRO infiltrates another pool and perform an attack. In such game, the honest CRO can control the payoffs and increase its revenue. By simulating this game, we propose to apply Zero Determinant (ZD) strategy on strategy decision, which can be categorized into cooperation and defecting. Our experimental results demonstrate the effectiveness of the proposed strategy decision method.*

## 1. INTRODUCTION

With the development of the Internet, cybersecurity becomes more and more important and serious. In the first half of 2018, according to the report published by CNVVD [1], 10,644 vulnerabilities were discovered, which exceeded 9,690 vulnerabilities in the same period of 2017. To protect cybersecurity, the users tend to seek professional security detection services, which are provided by a centralized trust third part. However, this mode has the following drawback. First, traditional security management is built in the centre environment, while attacks on the central nodes may devastate private data. Second, the traditional methods cannot provide a trust security platform for all participants, which can protect their privacy and avoid information leakage. Third, the white hat hackers earn little such that they almost have no interests.

To deal with the aforementioned problems, the blockchain-based security methods have been proposed. J.H et al. [3] study how to adapt blockchain security to cloud computing. In [4], blockchain-based approaches which improve the security of the Internet of Things (IoT) have been proposed. Blockchain can provide good solutions for security management and data storage [5-10].

Recently, a new blockchain-based security framework, Sapiens Chain, which can provide a trust vulnerability crowd testing environment and intelligent security detection services, was proposed. Sapiens Chain runs smart contracts on the blockchain, which guarantees the trustworthy and not be tampered for transactions. It schedules the CROs dynamically, where

CROs donate their own computing resources and are awarded after they finish the specific tasks. In order to increase the possibilities of rewarding, CROs tend to choose an open pool and cooperate with other CROs. In an open pool, CROs can be categorised into honest CROs and dishonest CROs. Dishonest CROs may reduce their consumption by forging work proofs, where they can also earn a certain amount of income with negative absenteeism. This is unfair for honest CROs, and thus we propose to tackle this challenge in this paper.

In this paper, we propose to use ZD (Zero-Determinant) strategies for CRO's selection. The ZD strategy was proposed in the Press and Dyson [15]. In the process of iterated games, one can use the ZD strategy to control the opponent's payoff unilaterally, so that the opponent's payoff maintains a linear relationship with hers. In other words, using ZD strategy can control the game unilaterally. As a probabilistic and conditional strategy, the ZD strategy has been widely employed in the iterated game, which aims to cope with the "free-riding" problems [11-13]. For example, Eyal et al. [14] qualitatively analysed the prisoner's dilemma in the mining process, which is a real instance of "free-riding" problems. Press et al. [15] proved that with ZD strategies, the player is able to unilaterally set the expected utility of an opponent or a ratio of the player's expected payoff to its opponent's, ignoring the opponent's strategy. Many other studies [16-21] on the strategy have shown the honest miners of blockchain can control opponents' payoffs.

In summary, we make the following contributions.

First, we review the overview of Sapiens Chain, including the architecture and the roles. Then we define the game that a CRO infiltrates another and perform an attack. In such kind of game, the honest CRO can control the payoffs and increase its revenue.

Second, we introduce different strategies for CROs and analyze how to apply Zero Determinant (ZD) on strategy decision, which can be categorized into cooperation or not.

Third, we report an extensive experimental study with numerical simulation. The results clearly show that the proposed strategy decision method is effective.

The rest of this paper is organized as follows. We review the overview of Sapiens Chain in Section 2. In Section 3, we analyze the CRO selection strategy in Sapiens Chain based on the ZD strategy. We report the empirical evaluation results in Section 4, and conclude the paper in Section 5.

## 2. THE OVERVIEW OF SAPIENS CHAIN

Sapiens Chain is a decentralized security detection platform, including the decentralized vulnerability platform and the automatic vulnerability detection system. It contains two kinds of nodes which are the ordinary nodes and fog nodes respectively.

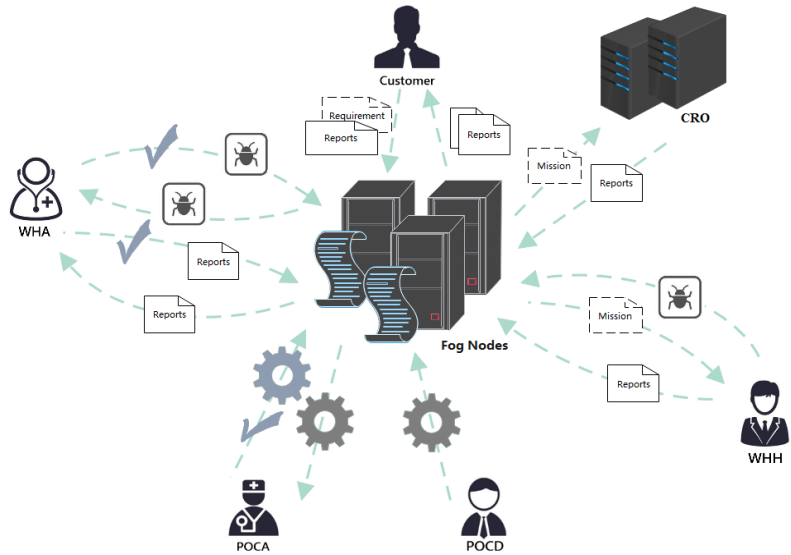

Figure 1. The architecture of Sapiens Chain

The ordinary nodes perform task publishing, POC writing, auditing and vulnerability detection, while the fog nodes are responsible for node scheduling, vulnerability storage, POC storage, vulnerability detection report storage, and key assignment for each ordinary node. There are 6 different roles on the ordinary nodes, which are the user, the POCD, the POCA, the WHH, the WHA and the CRO. The user submits the requirement, which needs to be detected by Sapiens Chain. The POCD provides POCs, which will be audited by the POCA. The WHH is the provider that can supply manual detection services, and the WHA tests the auditing reports submitted by WHHs.

The most important role in automatic detection is CRO, which is the computational resource provider. CROs in Sapiens Chain are similar to miners in blockchain. It deploys automated vulnerability detection tools and POCs, meanwhile being scheduled by the fog nodes with a pre-defined scheduling algorithm. As a participant in each mission, the CRO contributes its own computing power and then gains the benefit, which saves operating costs and makes Sapiens Chain decentralized. This model of blockchain also enhances the enthusiasm of the CRO and makes Sapiens Chain more robust.

The architecture of Sapiens Chain is shown in Figure 1. The user issues a requirement to the fog node, and the fog node schedules CROs, WHHs or WHAs according to the given requirements. CROs receive security plug-ins which are provided by POCDs and audited by POCAs and then run detection processes. The vulnerabilities submitted by WHHs will be audited by WHAs and finally received by the fog nodes, while WHHs can also submit vulnerabilities to users through the fog nodes. The fog nodes can distribute requirements and collect reports.

## 3. ANALYSIS ON STRATEGY DECISION FOR CROS

In this section, we first introduce the game between honest and dishonest CROs, and then propose the selection strategies in Sapiens Chain.

### 3.1 The Game model

The definitions of the honest CROs and dishonest CROs root in the Bitcoin mining problem. The dishonest CROs are also called attacker because they may initiate Block Withholding Attacks. That is, once the attackers have found the complete Proof-of-Work, they can choose to abandon the proof and only send partial Proof-of-Work to the mining pool. Through this kind of attack, the attackers will gain the payoff from the mining pool, but the mining pool can't benefit from the computing power of attackers. This will reduce the payoff of all CROs in this mining pool.

Sapiens Chain allows participants to use mining pools for increasing the computing power. Unfortunately, in Sapiens Chain mining pools, dishonest CROs will earn benefit through sabotage, such as counterfeiting Proof-of-Work. It is similar to the prisoner's dilemma which is a classic model of game theory and was first proposed by Albert Tucker [22].

The prisoner's dilemma describes the game process of two prisoners, which is shown as follows. The police separately interrogate two prisoners to prevent collusion. At the same time, the police offers several options, that is, (1) if X and Y both remain silent, both of them will only serve 3 years in prison; (2) if X betrays Y but Y remains silent, X will be set free and Y will serve 5 years in prison; (3) if X and Y both betray the other, each of them will serves 1 year in prison.

Table 1. The payoff matrix in prisoner's dilemma.

|   | C      | D      |
|---|--------|--------|
| C | (R, R) | (S, T) |
| D | (T, S) | (P, P) |

Similar to the above situation, we introduce the prisoner's dilemma in Sapiens Chain as follows. Denote C, D, R, S, T and P as "Cooperation", "Defecting", "Reward for mutual cooperation", "Sucker's payoff", "Temptation to defect" and "Punishment for mutual defection" respectively. Table 1 shows the prisoner's dilemma payoff matrix in Sapiens Chain. Since T>R>P>S, the prisoner's dilemma game in Sapiens Chain exists. Dishonest CROs choose to attack for increasing benefits, resulting in a decreasing of honest CROs' payoff. Facing this prisoner's dilemma, the CROs tend to raise their own payoff by launching attacks.

### 3.2 ZD strategies in Sapiens Chain

In the open pool of Sapiens Chain, the interaction between honest CROs and dishonest CROs is regarded as an iterative game. At each iteration, we model the process as a single stage of prisoner's dilemma game and we suppose that long-memory player has no advantage over the short-memory player. Thus, the actions of both CROs only depend on the rewards obtained from the previous round.

Let $\boldsymbol{p} = (p_1, p_2, p_3, p_4)$ and $\boldsymbol{q} = (q_1, q_2, q_3, q_4)$ be the probabilities of cooperation or attacking based on the previous iteration for the honest CRO X and the dishonest CRO Y. We use a parameter $m(0 < m < 1)$ to represent the reduced ratio of the cooperation rate for CROs who betrayed last time. Thus we can rewrite $p$ and $q$ as $\boldsymbol{p} = (p_1, p_2, mp_3, mp_4)$ and $\boldsymbol{q} = (q_1, mq_2, q_3, mq_4)$ respectively, which represent the transition probability vectors for the cooperation state in the next round. Then we mode an iterated game as a Markov process, and we can build the Markov matrix as shown in Equation (1).

$$P = \begin{bmatrix} p_1 q_1 & p_1(1-q_1) & (1-p_1)q_1 & (1-p_1)(1-q_1) \\ p_2 m q_3 & p_2(1-m q_3) & (1-p_2)m q_3 & (1-p_2)(1-m q_3) \\ m p_3 q_2 & m p_3(1-q_2) & (1-m p_3)q_2 & (1-m p_3)(1-q_2) \\ m p_4 m q_4 & m p_4(1-m q_4) & (1-m p_4)m q_4 & (1-m p_4)(1-m q_4) \end{bmatrix} \quad (1)$$

Since matrix $P$ has a unit eigenvalue, the matrix $M = P - I$ (see Equation (2)) must be a singular matrix and $|M| = 0$. $M$ can be expressed as:

$$M = \begin{bmatrix} p_1 q_1 - 1 & p_1(1-q_1) & (1-p_1)q_1 & (1-p_1)(1-q_1) \\ p_2 q_3 & p_2(1-q_3)-1 & (1-p_2)q_3 & (1-p_2)(1-q_3) \\ m p_3 m q_2 & m p_3(1-m q_2) & (1-m p_3)m q_2 - 1 & (1-m p_3)(1-m q_2) \\ m p_4 m q_4 & m p_4(1-m q_4) & (1-m p_4)m q_4 & (1-m p_4)(1-m q_4)-1 \end{bmatrix} \quad (2)$$

Let $v^T P = v^T$, $v^T M = 0$. Then we can obtain the stationary vector $v = [v_1, v_2, v_3, v_4]^T$ of $M$. Given $M^*$ as shown in Equation (3), according to the Cramer's ruler, we have $M^* M = \det(M) I = 0$, where $c_{ij} = (-1)^{i+j} \det(M'_{ij})$ and $[v_1, v_2, v_3, v_4]$ is proportional to $[c_{14}, c_{24}, c_{34}, c_{44}]$. It follows,

$$v^T f = [v_1, v_2, v_3, v_4] \begin{bmatrix} f_1 \\ f_2 \\ f_3 \\ f_4 \end{bmatrix} \equiv D(p, q, f)$$

$$= \det \begin{bmatrix} -1 + p_1 q_1 & -1 + p_1 & -1 + q_1 & f_1 \\ p_2 m q_3 & -1 + p_2 & m q_3 & f_2 \\ m p_3 q_2 & m p_3 & -1 + q_2 & f_3 \\ m p_4 m q_4 & m p_4 & m q_4 & f_4 \end{bmatrix} \quad (3)$$

Let $f = \alpha S_X + \beta S_Y + \gamma 1$. The expected payoff of CRO X and CRO Y in the stationary state satisfies Equation 4, where $\alpha, \beta, \gamma$ are non-zero parameters.

$$\alpha S_X + \beta S_Y + \gamma 1 = \frac{D(p, q, \alpha S_X + \beta S_Y + \gamma 1)}{D(p, q, 1)} \quad (4)$$

Let $p = (p_1 - 1, p_2 - 1, m p_3, m p_4)^T = \varphi(\alpha S_X + \beta S_Y + \gamma 1)$. It follows

$$p_1 = 1 + \varphi\left((\alpha+\beta)\frac{r-c}{2}+\gamma\right)$$

$$p_2 = 1 + \varphi\left(\alpha\left(\frac{r}{2}-c\right)+\beta\frac{r}{2}+\gamma\right)$$

$$p_3 = \frac{\varphi\left(\beta\left(\frac{r}{2}-c\right)+\alpha\frac{r}{2}+\gamma\right)}{m}$$

$$p_4 = \frac{\varphi\gamma}{m}$$

(5)

From the above analysis, we can know that if the CRO X takes strategies of $\alpha S_X$, $\beta S_Y$ or a linear combination of them, it will have the possibility to choose unilateral strategies and make the determinant in numerator in Equation (4) vanish. Similarly, the same to CRO Y. Thus if the CRO X adopts a strategy which satisfies $\boldsymbol{p} = \alpha \boldsymbol{S}_X + \beta \boldsymbol{S}_Y + \gamma \boldsymbol{1}$, or if the CRO Y takes a strategy with $\boldsymbol{q} = \alpha \boldsymbol{S}_X + \beta \boldsymbol{S}_Y + \gamma \boldsymbol{1}$, then the determinant vanishes. Accordingly, the payoffs of CRO X and CRO Y will be satisfied with a linear relationship

$$\alpha \boldsymbol{S}_X + \beta \boldsymbol{S}_Y + \gamma \boldsymbol{1} = \boldsymbol{0}.$$

### 3.3 The ZD-set strategy in Sapiens Chain

In Sapiens Chain, the ZD-set strategy means that honest CRO X uses the ZD strategy to unilaterally set the long-term benefits of dishonest CRO Y. When honest CRO X uses the ZD-set strategy, we fix $\alpha = 0$ in $\alpha \boldsymbol{S}_X + \beta \boldsymbol{S}_Y + \gamma \boldsymbol{1} = \boldsymbol{0}$. Let $\tilde{\boldsymbol{p}} = \beta \boldsymbol{S}_Y + \gamma \boldsymbol{1}$. We can use $p_1$ and $p_4$ to represent $p_2, p_3$ and $\boldsymbol{S}_Y$, which are shown in Equation (6)-(8).

$$p_2 = \frac{rp_1 - c(1+p_4)}{r-c} \tag{6}$$

$$p_3 = \frac{(2c-r)(1-p_1)+cp_4}{r-c} \tag{7}$$

$$S_Y = \frac{p_4(r-c)}{2(1-p_1+p_4)} \tag{8}$$

Thus, we can conclude that, when CRO X takes a ZD-set strategy, it can unilaterally set the payoff of CRO Y, regardless of the CRO Y's strategy, while the CRO X can't control its own payoff even with any subclass of ZD strategy.

### 3.4 The ZD-extortion strategy in Sapiens Chain

In Sapiens Chain, the ZD-extortion strategy means that honest CRO X uses the ZD strategy to unilaterally set the long-term return of dishonest CRO Y to be linear with his own earnings. If honest CRO X uses the ZD-extortion strategy, it can also guarantee that the other party's payoff is always lower than its own payoff. Since ZD-extortion strategy always passes the reference point $l = P$, we have $\tilde{p} = \varphi(s(S_X - P) - (S_Y - P))$, where $s < 1$. $p_1, p_2, p_3, p_4$ can be represented as follows, which are shown in Equation (9).

$$\begin{aligned} p_1 &= 1 - \varphi(1-s) \cdot \frac{r-c}{2} \\ p_2 &= 1 - \varphi\left(s\left(c - \frac{r}{2}\right) + \frac{r}{2}\right) \\ p_3 &= \varphi\left(\left(c - \frac{r}{2}\right) + s \cdot \frac{r}{2}\right) \\ p_4 &= 0 \end{aligned} \quad (9)$$

If $\varphi$ is small enough, there exists a viable extortion strategy for any $s$. Since $p_1, p_2, p_3, p_4$ are all between $[0,1]$. It follows,

$$\begin{aligned} 0 &\leq 1 - \varphi(1-s) \cdot \frac{r-c}{2} \leq 1 \\ 0 &\leq 1 - \varphi\left(s\left(c - \frac{r}{2}\right) + \frac{r}{2}\right) \leq 1 \\ 0 &\leq \varphi\left(\left(c - \frac{r}{2}\right) + s \cdot \frac{r}{2}\right) \leq 1 \end{aligned} \quad (10)$$

The range of $\varphi$ can be represented in Equation (11).

$$0 \leq \varphi \leq \frac{1}{s\left(c - \frac{r}{2}\right) + \frac{r}{2}} \quad (11)$$

Since $\varphi$ is small enough, the range of extortion factor s can satisfy $\frac{r-2c}{r} \leq s < 1$. Following ZD-extortion strategy, CRO X will have a larger payoff than CRO Y.

## 4. EXPERIMENTS

In this section, we use numerical simulation to illustrate the performance of the different strategies in the prisoner's dilemma. We use WSLS, ALLD and ALLC as the comparison strategies, where WSLS strategy is "win stay, lose shift", ALLD strategy is "Always Defecting" and ALLC strategy is "Always Cooperation". Fix $R = 1.5$, $S = -1$, $T = 3$, $P = 0$ then the payoff vector of the honest CRO X will be $S_X = (1.5, -1, 3, 0)$, and that of the dishonest CRO Y will be $S_Y = (1.5, 3, -1, 0)$. For each experiment, we evaluate the differences of the payoffs between the honest CRO X and the dishonest CRO Y. All figures show the possible payoffs of the CRO X (on the horizontal axis) and the CRO Y (on the vertical axis) as colored areas or lines, where the colored points represent the payoff pairs for 50, 000 chosen opponents.

In Figure 2, when honest CRO X adopts the WSLS strategy and dishonest CRO Y takes a strategy randomly, the payoff of the two CROs shows that the profit coverage area is a triangle.

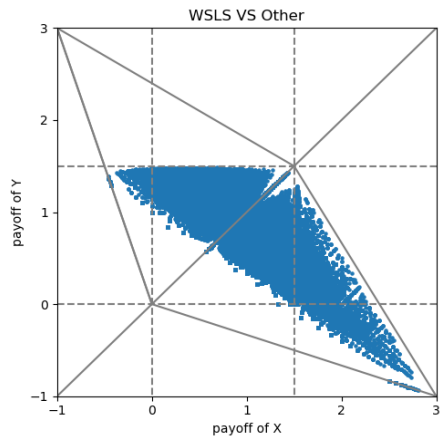

Figure 2. The payoff of two CROs (WSLS VS Random)

In Figure 3, when honest CRO X always adopts a cooperative or betrayal strategy, and dishonest CRO Y takes a strategy randomly, the payoffs of the two CROs shows that the profit coverage area is a straight line.

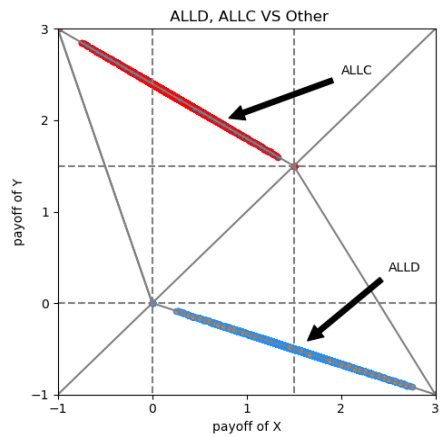

Figure 3. The payoff of two CROs (ALLC, ALLD VS Random)

In Figure 4, when honest CRO X adopts the ZD-set strategy and dishonest CRO Y also takes a strategy randomly, the profit of the two CROs shows that the honest CRO X can dominate the opponent's benefit which is always on a straight line.

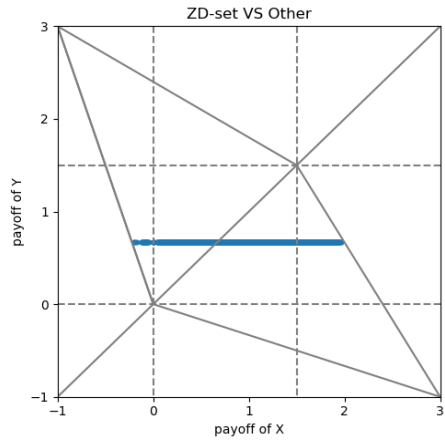

Figure 4. The payoff of two CROs (ZD-set VS Random)

In Figure 5, when honest CRO X adopts the ZD-extortion strategy and dishonest CRO Y takes a strategy randomly, the payoffs of the two CROs shows that the CRO X can control not only the opponent's income which is always on a straight line, but also his benefit higher than s times the other's.

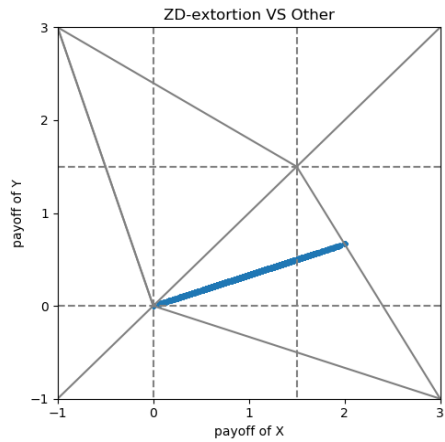

Figure 5. The payoff of two CROs (ZD-extortion VS Random)

Compared with ZD strategy, the WSLS, ALLC, and ALLD strategies have no restrictions on the payoffs. When it comes to the ZD-set strategy, it can achieve the goal that honest CRO X restricts dishonest CRO Y, and the ZD-extortion strategy can guarantee that the profit of honest CRO X is higher than dishonest CRO Y's. We attempt to apply ZD strategy to Sapiens Chain. No matter what strategy the dishonest CROs adopt, when honest CROs adopt the ZD strategy, the payoff of dishonest CROs can be restricted. What's more, honest CROs can keep the payoffs of dishonest CROs linear with their own earnings, which makes it possible to design efficient game consensus.

## 5. CONCLUSION

In this paper, we first define a game that a CRO infiltrates another and perform an attack. In such game, the honest CRO can control the payoff and increase its revenue. By simulating this game, we propose to apply Zero Determinant (ZD) on strategy decision, which can be categorized into cooperation or not. Our experimental results demonstrate the effectiveness of the proposed strategy decision method, indicating that the honest CROs can apply ZD strategy to control the payoffs of themselves higher than dishonest CROs. Due to ZD strategy, CROs become more energetic and Sapiens Chain becomes safer.